\def\ion#1#2{#1$\;${\small\rm\@Roman{#2}}\relax}

\documentclass[rnote]{aa}

\usepackage{graphicx}
\usepackage{txfonts}
\newcommand{\aanda}  {A\&A \nolinebreak}

\newcommand{\etal}    {et al.}
\usepackage{natbib}
\begin{document}
   \title{Examining the evidence for dust destruction in GRB\,980703}
   \author{Rhaana~L.~C.~Starling
          }

   \offprints{R.~L.~C.~Starling}

   \institute{Dept. of Physics and Astronomy, University of Leicester, Leicester LE1 7RH, UK\\
              \email rlcs1@star.le.ac.uk}

   \date{Received ; accepted }

   \abstract{{The effects that gamma-ray bursts have on their environments is an
       important and outstanding issue. Dust destruction in particular has
       long been predicted while observational evidence is difficult to
       obtain. We examine the evidence for dust destruction by GRB\,980703, in
       which various inconsistent measurements of the host galaxy extinction have been made
       using the GRB afterglow emission.}{ We construct a spectral energy
       distribution from nIR to X-ray to measure the extinction at early times
     and compare this with previous findings. We also construct nIR/optical
     SEDs at intermediate epochs to examine a previously reported decrease in extinction.}{ The extinction is very
     high for a GRB host galaxy. The earliest extinction measurement is likely to be
     lower than previously estimated, and consistent with most later
     measurements. In a series of SEDs we do not find any evidence of
     variable extinction. We therefore conclude that there is no clear evidence of
     dust destruction in this case.}{}
   }

   \titlerunning{Examining the evidence for dust destruction in GRB\,980703}
   \authorrunning{R.L.C. Starling} 
  \maketitle
%

\section{Introduction}
The search for the predicted effects of gamma-ray bursts (GRBs) on their environments is
key to understanding the properties of their host galaxies, found both nearby
and in the early Universe. Ionisation by the powerful GRB and afterglow emission has now been seen in
the form of optical absorption line variability (e.g. GRB\,020813, Dessauges-Zavadsky et
al., 2006; GRB\,060418, Vreeswijk et al., 2007) and hinted at from several reports of low significance
variations in X-ray column density for some GRBs, for example GRB\,011121 \citep{piro}, GRB\,050730
\citep{050730}, GRB 050904 \citep{boer,campana,gendre2} and GRB 060729 \citep{grupe} and from
comparison of optical and X-ray hydrogen column densities \citep{watson}. 

Optical extinction should
also decrease as we expect dust to be destroyed by the GRB jet out to 10--30 pc
\citep{drainesal,waxdraine,fruchter,perna}. This should occur very shortly
after the onset of the GRB, but so far only one such GRB had potentially shown
this effect, GRB\,980703, and the decrease in extinction would have occurred
from approximately 1 day after the GRB onwards \citep[e.g.][]{vreeswijk1,holland,columnsI}. The evidence of this comes from two sources. Firstly the
suggestion by \cite{castrotirado} that the spectral slope estimated from
their $R$- and $H$-band photometric data from around 1 day after the burst indicated a redder
value than would normally be expected for GRBs, translating into $A_V$ $\sim$
2.2. This is large when compared with the average value of $A_V= 1.07$
found from afterglow studies at later
times. Secondly, \cite{vreeswijk1} compiled spectral energy
distributions (SEDs) at four epochs covering 2.2--5.2 days after the burst and derive a potential decrease in the intrinsic
optical extinction at the 2--3$\sigma$ level. 

\begin{figure}
   \centering
   \includegraphics[angle=90,width=8cm]{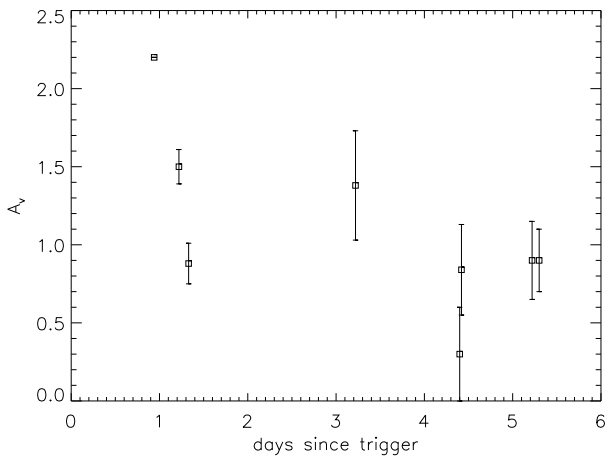}
   \vspace{0.6cm}
   \caption{Optical extinction ($A_V$) values reported in the literature
     (squares). All errors
   are plotted at the 1$\sigma$ confidence level.}         
\label{literature}
   \end{figure}

There have been optical extinction measurements
of GRB\,980703 by five groups using a number of different methods and instrumentation, and measurements of $A_{V}$ at different
times are far from consistent (Figure \ref{literature}; Table \ref{tab:literature}).
There is, therefore, a need to return to the apparent variability of
$A_V$ in GRB\,980703, which if confirmed would be the first observational
evidence of this kind. We do this
via broadband spectral energy distributions, where possible combining an X-ray spectrum with
early near-infrared (nIR) and optical data, and thereby discuss
the observational manifestations of dust destruction by GRBs. 
This issue is
particularly important to re-examine in the current era of rapid follow-up by
the {\it Swift} satellite \citep{gehrels}, robotic telescopes
(e.g. Liverpool Telescope and Faulkes Telescopes, 2m in diameter) and rapid
response mode on larger ground-based telescopes such as the 4m William
Herschel Telescope and the 8m Very Large Telescopes. 

\begin{table*}
\begin{center}
\renewcommand{\arraystretch}{2.0}
\begin{tabular}{|cc|ll|} \hline
Time since burst (days) & $A_V$ & Method & Reference \\ \hline
0.94 & 2.2 & nIR/optical colour & Castro-Tirado et al. 1999\\
1.22 & 1.5$\pm$0.1 & nIR/optical/X-ray SED continuum & Vreeswijk et al. 1999 \\
1.33 & 0.9$\pm$0.1 & nIR/optical/X-ray SED continuum & Starling et al. 2007 \\
3.22& 1.38$\pm$0.35 & nIR/optical SED continuum & Vreeswijk et al. 1999 \\
4.4 & 0.3$\pm$0.3 & host spectrum Balmer decrement\footnotemark & Djorgovski et al. 1998\\
4.42& 0.84$\pm$0.29 & nIR/optical SED continuum & Vreeswijk et al. 1999 \\
5.22& 0.90$\pm$0.25 & nIR/optical SED continuum & Vreeswijk et al. 1999 \\
5.3 & 0.9$\pm$0.2 & nIR/optical SED continuum & Bloom et al. 1998 \\ \hline
\end{tabular} 
\caption{Measurements of optical extinction gathered from the
  literature. Errors are either given at the 1$\sigma$ level or assumed to be so where no confidence level is stated.}
\label{tab:literature}
\end{center}
\end{table*}

\section{Re-evaluating the earliest $A_V$ estimate}
\subsection{Method}
We use a broadband dataset to create a spectral energy distribution at early
times when a high value of $A_V$ is reported from optical data.
We took the earliest magnitudes reported in \cite{castrotirado} ($R$ and $H$ band) which
were taken at 0.94 days since burst (and used for their estimates of $A_V$). 
We converted the observed magnitudes into fluxes
assuming the Johnson filters, after applying a correction for the Galactic
extinction of $E(B-V)=0.057$ \citep{schlegel}. The Lyman-$\alpha$ Forest
correction \citep[e.g.][]{madau},
due to the redshift of $z=0.9661$ \citep{djorg,vreeswijk1}, is
negligible for these red bands. 

We then calculated the required end-time for
the {\it BeppoSAX} \citep{boella} X-ray observations in order to obtain a log observation mid-time of 0.94
days, given that the observations began on 0.827 days (Low Energy Concentrator
Spectrometer, LECS, covering $0.1-4$ keV) and 0.844 days
(Medium Energy Concentrator Spectrometer, MECS, covering $1.4-10$ keV). We extracted a MECS spectrum with these time intervals (up to 0.96
days) and found that only of order 80 counts were present. This did not allow
us to create a spectrum of sufficient quality.
The observed $2-10$ keV X-ray flux at 1 day is derived from the light curve in \cite{gendre1} to be (4.8$\pm$0.07) $\times$10$^{-13}$ erg cm$^{-2}$ s$^{-1}$ with a
decay index of $\alpha_X$ = 0.9$\pm$0.2, (where $F_{\nu} \propto
t^{-\alpha}\nu^{-\beta} \propto t^{-\alpha}\nu^{-(\Gamma-1)}$). We note that the
early optical decay index of $\alpha_O$ = 0.85 $\pm$ 0.84 given in \cite{zeh}
is consistent with this (though note the large associated error).
We used this information to
transform the full {\it BeppoSAX} LECS and MECS X-ray spectra
presented in \cite{columnsI} (log
observation mid-time 1.33 days) to the time of the early optical
photometry. 

We fitted the optical and X-ray data together in a count space
fit following the method outlined in \cite{columnsI}. 
The underlying continuum is fit with an absorbed power law. Previously, when taking into account
consistency with the fireball model \citep{fireball1,fireball2,fireball3,fireball4,fireball5}, several authors could estimate the position of the cooling
break, $\nu_{\rm c}$. This break appears to move to lower frequencies with
time \citep{vreeswijk1,bloom,columnsII}.
It follows from these
results that at 0.9 days since burst we expect a single power law shaped
spectrum spanning the nIR/optical to X-ray regime. Galactic absorption is
taken into account using $N_{\rm H} = 4.98\times 10^{20}$ cm$^{-2}$
\citep{kalberla}. Galactic extinction is fixed at $E(B-V) = 0.057$ \citep{schlegel}.
We modelled the extinction
in the nIR/optical bands using the three most well-known extinction curves of
the Milky Way (MW), Large Magellanic Cloud (LMC) and Small Magellanic Cloud (SMC), as parametrised by \cite{pei}.
\footnotetext{We note that
  this method probes a different volume than the other, line-of-sight, methods
  listed and should therefore not be directly compared. For a discussion on
  methods of estimating extinction see e.g. \cite{mathis}, \cite{savaglio} and \cite{columnsI}.}
\subsection{Results}
\begin{table}
\begin{center}
\renewcommand{\arraystretch}{2.0}
\begin{tabular}{|l|ccc|c|} \hline
Model & $\Gamma_1$ & $A_V$ & $N_{\rm H}$ &
$\chi^2$/dof \\
 & &  & $\times$10$^{22}$ cm$^{-2}$ & \\ \hline
PL+MW & 2.05$\pm$0.02 & 1.24$^{+0.10}_{-0.12}$ & 0.8$^{+0.6}_{-0.4}$
& 32/24\\
PL+LMC & 2.05$\pm$0.02 & 1.14$\pm$0.10 & 0.8$^{+0.6}_{-0.4}$ & 32/24\\
PL+SMC & 2.04$\pm$0.02 & 1.08$^{+0.11}_{-0.09}$ &
0.7$^{+0.6}_{-0.4}$ & 33/24\\ \hline
\end{tabular}
\caption{Absorbed power law fits to the SED at 0.9 days since burst. All
  errors are quoted at the 1$\sigma$ confidence level.}
\label{tab:sed}
\end{center}
\end{table}
The results of SED fits are given in Table \ref{tab:sed}. A single absorbed power law is a statistically acceptable fit to these data
with power law photon index $\Gamma\sim2.0$, X-ray
absorption of order 7$\times$10$^{21}$ cm$^{-2}$ and optical extinction of
$E(B-V)\sim0.4$ (approximately equivalent to $A_V \sim 1$).
Comparing Table \ref{tab:sed} with Table \ref{tab:literature} shows that the extinction
values are consistent at the 1$\sigma$ level
with those derived from fits at 3.22, 4.42 and 5.22 days since burst \citep{vreeswijk1},
1.33 days since burst \citep{columnsI} and at 5.3 days since burst
\citep{bloom}. At this confidence level the
result is inconsistent with previous estimates at 0.94 days \citep{castrotirado}, 1.22 days \citep{vreeswijk1} and 4.4 days \citep[][host galaxy measurement]{djorg}.   
MW, LMC and SMC extinction curves are indistinguishable for these data; this is partially because we do not have data in the bluer bands where the 2175\AA\
bump characteristic of only the MW curve would occur at this redshift. The SMC
extinction curve is generally the best approximation to GRB host galaxy extinction
laws of the three aforementioned curves, as shown in many studies and most
likely following the low metallicity pattern of GRB hosts \citep[e.g.][]{galwij,stratta,kann,schady,columnsI}.

 \begin{figure}
   \centering
   \includegraphics[angle=-90,width=8cm]{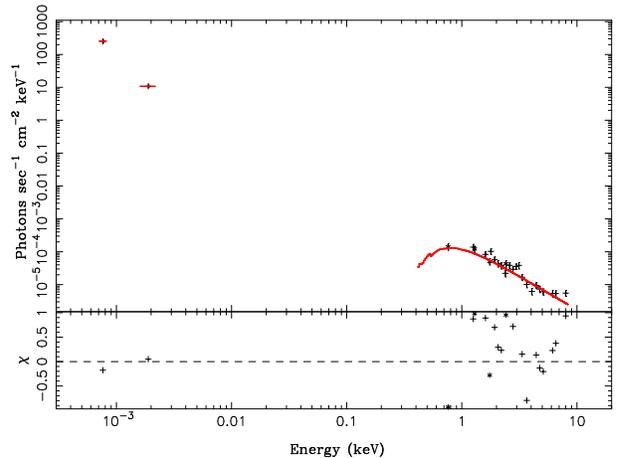}
   \caption{Unfolded SED (instrumental response at X-ray energies removed) at 0.9 days since burst (upper panel) and residuals to the fit
     (lower panel). The black crosses (+) are the optical
     photometry and MECS spectrum and the black stars show the LECS
     spectrum. The solid (red) lines are the model of a power law absorbed in X-rays and
   in optical by an SMC-like extinction.}         
\label{seds}
   \end{figure}

\section{Re-evaluating variability claims}
\subsection{Method}
At four later epochs spanning $1.2-5.2$ days since burst \cite{vreeswijk1} created nIR/optical SEDs
and find a decrease in the intrinsic optical extinction of $\Delta A_V = 0.24-0.96$. 
In order to test the possible decreasing extinction proposed by these authors we recreated the nIR/optical SEDs at the epochs 3.2, 4.4 and
5.2 days since trigger. We do not reproduce the SED at 1.2 days because this epoch (optical data from 1.2 days extrapolated to 1.3
days) has been covered using the same SED
method and models in \cite{columnsI} and we will use the results from that
fit. Likewise, the previously reported epochs of 5.2 and 5.3 days
\citep[][respectively]{vreeswijk1,bloom} are close in time so are treated here
with a single SED.
We gathered all available data \citep[][and references
therein]{vreeswijk1,castrotirado} to create light curves in seven photometric
bands, $BVRIJHK$ (the $B$ band was not included in the SEDs of \cite{vreeswijk1}), interpolating from the six nearest points in time to find
the magnitude per epoch. We corrected for the host galaxy contribution to each
band using the host magnitudes from \cite{sokolov} ($B$ band) and \cite{vreeswijk1} ($VRIJHK$) and errors on the host magnitude determinations were combined with the datapoint and fit errors.

\subsection{Results} 
\begin{figure}
   \centering
   \includegraphics[angle=0,width=8cm]{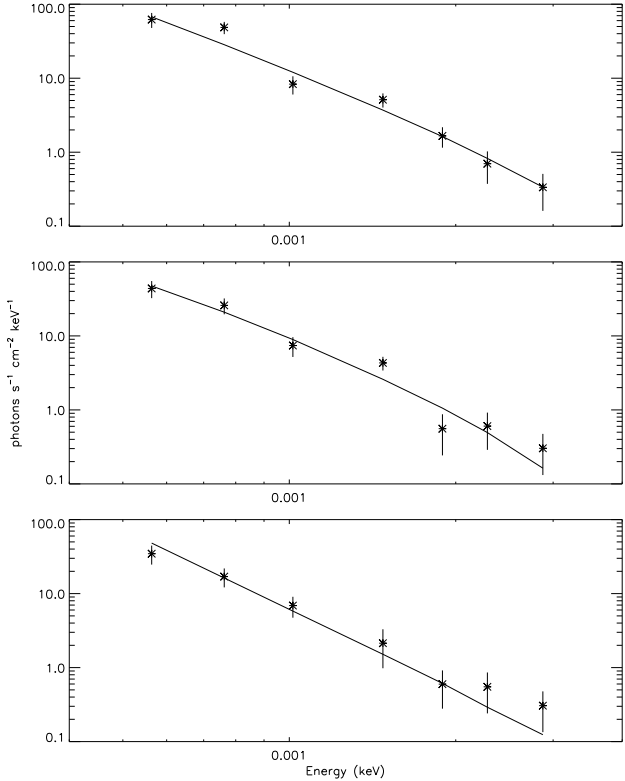}
   \caption{nIR/optical SEDs at 3.2 (upper), 4.2 (middle) and 5.2 (lower panel)
     days since burst with best-fitting power law model plus SMC extinction
     overlaid (solid lines). The SEDs are plotted in the same units as the
     optical--X-ray SED of Fig. \ref{seds} for comparison.}
\label{newseds}
   \end{figure}

The SEDs are shown in Fig. \ref{newseds} and fit results are presented in Table \ref{tab:newseds}. 
Fitting a model consisting of a power law plus fixed Galactic extinction and variable (SMC- or MW-like) host extinction, we find we cannot constrain both the power law slope
and the host extinction simultaneously. We derive upper limits on $A_V$ which are consistent with each other for these three epochs, and with most previous
measurements.
\begin{table}
\begin{center}
\renewcommand{\arraystretch}{2.0}
\begin{tabular}{|l|l|cc|c|} \hline
Epoch &Model & $\Gamma$ & $A_v$ & $\chi^2$/dof \\ \hline
3.22 &PL+SMC&2.6$^{+0.5}_{-0.9}$&$<$1.1 & 9.4/4 \\
     &PL+MW&2.7$^{+0.4}_{-0.9}$&$<$1.2 & 9.4/4 \\
     &PL+SMC&2.013&0.82$^{+0.23}_{-0.18}$ & 9.7/5 \\
     &PL+MW&2.013&0.87$^{+0.25}_{-0.19}$ & 10.2/5 \\
     &PL+SMC&1.87&0.94$^{+0.23}_{-0.18}$&9.9/5 \\
     &PL+MW&1.88&0.99$^{+0.25}_{-0.22}$&10.4/5 \\ \hline
4.42 &PL+SMC&2.1$^{+0.9}_{-0.8}$&$<$2.2&8.1/4 \\
     &PL+MW&3.0$\pm$0.3&$<$2.3&8.7/4 \\
     &PL+SMC&2.013&0.97$^{+0.26}_{-0.23}$&8.1/5 \\
     &PL+MW&2.013&1.05$^{+0.28}_{-0.25}$&9.2/5 \\
     &PL+SMC&1.87&1.08$^{+0.26}_{-0.23}$&8.1/5 \\
     &PL+MW&1.88&1.18$\pm$0.28&9.3/5 \\ \hline
5.22 &PL+SMC&3.0$\pm$0.4&$<$1.8&1.5/4 \\
     &PL+MW&3.0$^{+0.3}_{-0.4}$&$<$1.9&1.55/4 \\
     &PL+SMC&2.013&0.88$^{+0.35}_{-0.26}$&1.9/5 \\
     &PL+MW&2.013&0.96$^{+0.40}_{-0.31}$&2.7/5 \\
     &PL+SMC&1.87&1.00$^{+0.38}_{-0.26}$&1.99/5 \\
     &PL+MW&1.88&1.09$^{+0.40}_{-0.34}$&2.8/5 \\ \hline
\end{tabular}
\caption{Absorbed power law fits to the SEDs at epochs 3.2, 4.4 and 5.2 days since
  burst. All
  errors and upper limits are quoted at the 1$\sigma$ confidence level.}
\label{tab:newseds}
\end{center}
\end{table}

We may hope to reproduce the decrease in $A_V$ if we keep the power
law slope fixed, as done by \cite{vreeswijk1}. They fixed $\beta_{\rm opt}=1.013$ (or
$\Gamma=2.013$) which they measured from a single power law fit to their
optical--X-ray SED at 1.2 days since trigger. At a similar time, 1.3 days, \cite{columnsI}
measured $\beta_{\rm opt}=0.87$ (SMC) and 0.88 (MW) (or $\Gamma=1.87,1.88$)
from a broken power law fit to an optical--X-ray SED, so we tested both fixed values for the power law slope.
Making this assumption, that the power law slope does not change in the optical
regime over this time period, we consistently find values of $A_V \sim 0.8-1.2$, in agreement with the values reported in the initial study for the
final two epochs but lower than those of the first two epochs
\citep{vreeswijk1}. Of those tested, no one extinction model is preferred.

\section{Discussion: is there evidence for dust destruction in GRB\,980703?}
Construction of a broadband SED at early times (0.9 days) results in a lower value for the optical extinction
than previously found using the optical colour, bringing this first extinction
estimate into line with all later afterglow measurements at the 2$\sigma$
level or better (Fig. \ref{literature}). We then addressed the possible decrease
in extinction claimed for several later epochs by creating new SEDs at these
times with both increased wavelength coverage and fitting with a wider range
of extinction models.
It appears that given the substantial errors on most afterglow-derived $A_V$
measurements they are all consistent when the SEDs are
recreated in a consistent manner and fit with the same models. We have tested
the need for decreasing extinction across all our newly derived $A_V$ measurements
spanning $0.9-5.2$ days since trigger by performing a linear fit to the five
values and their 1$\sigma$ errors. We adopted the SMC extinction curve in all
cases, with the power law photon index in the optical regime as measured at
0.9 and 1.2 days and fixed at 2.013 at 3.2, 4.4 and 5.2 days (all consistent with
$\Gamma=2.0$). Throughout this time period the optical spectral slope should be
well described by a single power law. The cooling break is proposed to lie in
the medium energy X-ray band early on (e.g. at $\sim$8$\times$10$^{17}$ Hz at 1.3 days
\citep{columnsII}) to account for the X-ray temporal slope within the fireball
model (hence a broken power law model was adopted for the broadband SED at
this time) and in the soft X-ray band or UV five days since trigger \citep{bloom}. 
In a homogeneous circumburst medium the cooling break would move to
lower frequencies with time approximately as t$^{0.5}$ up until the onset of
any jet break, and would remain within the X-ray band throughout
the epochs covered by this study.

\begin{figure}
   \centering
   \includegraphics[angle=90,width=8cm]{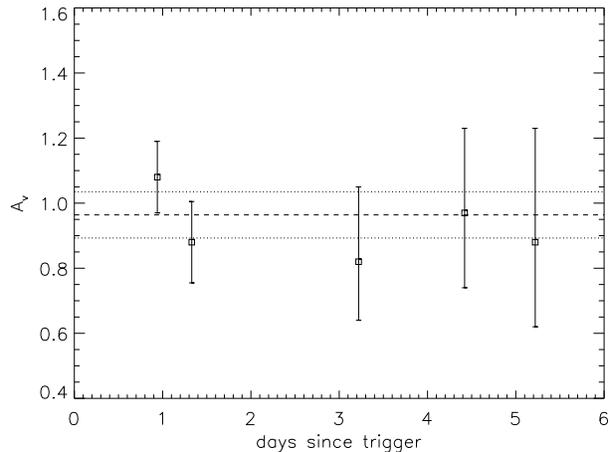}
   \caption{Optical extinction ($A_V$) values derived in this study from
       afterglow SED fits
     (squares). The dashed line shows the
   best fitting constant value to the distribution and the dotted lines show
   its associated error. All errors
   are plotted at the 1$\sigma$ confidence level.}         
\label{finalextincs}
   \end{figure}

We find that a single value of $A_V$ = 0.96$\pm$0.07 is an acceptable description of
the data, with a fit statistic of $\chi^2$/dof = 2.13/4. This is shown in
Figure \ref{finalextincs}. A fit to the data
using a polynomial of degree 1 (allowing for an increase or decrease in the extinction) returned a fit statistic of $\chi^2$/dof = 1.62/3, which does not
significantly improve the fit; the F-Test probability for inclusion of the additional free
parameter is 0.4, i.e. there is a 40\% chance probability of obtaining this
result.  
While the lack of broadband (X-ray and UV) data at these epochs prevents us from definitively ruling out variable $A_V$, the
results strongly suggest that a decreasing extinction is not
required. This may be due to a combination of the additional data used in our
study, application of the SMC extinction curve and a different fitting
procedure. We cannot identify the preferred extinction model in any of our fits, again because
more optical bands, particularly towards the UV where the curves deviate most
significantly, are needed. This also means that where the
extinction curves applied to estimate the previous $A_V$ values
differ, this is unlikely to affect the overall results. \cite{kann} have shown that it is difficult to
distinguish between optical extinction models for $z<1.5$ of which GRB\,980703 ($z=0.9661$) is an
example.

Most models predict some form of ionisation and dust destruction by the GRB on
its immediate surroundings. To observe this we need to probe distances of
perhaps up to
$10-30$ parsec from the GRB \citep{waxdraine}. 
Decreasing X-ray column densities, while not yet measured at a highly
significant level, suggest ionisation by the GRB is taking place. Comparison
of \ion{H}{I} column densities from optical spectra with equivalent hydrogen columns
measured from X-ray spectra concluded that the optical and X-ray gas are
likely not co-located, and the wide spread in measured values implies that
ionisation plays a role \citep{watson}. These authors estimate a maximum distance out to
which most of the hydrogen has been ionized by the GRB of 3 parsec.
Time resolved optical
spectroscopy has now revealed compelling evidence of UV pumping, seen as
variations in optical absorption line properties, most evident in GRB\,060418
\citep{vreeswijk2}. These results suggest that the optical regime is probing gas that lies
1.7 kiloparsec from the GRB site, further than the distance out to which dust is
thought to be destroyed. 

Dust destruction will also only take place as the GRB
ionising flux reaches the cloud, perhaps only minutes or seconds (in the
observer frame) after the GRB
has occurred depending on the distance to the cloud and the number and energy
of ionising photons, and should be temporally coincident with the ionisation. To return to the example of GRB\,060418, we note that
observations began 11 minutes after the burst. Changes in the X-ray column
density thought to be due to ionisation by the GRB are on similar
timescales, for example in GRB\,050730 this occurred over the first 8 minutes
since trigger. In the case of GRB\,980703
accurate localisation occurred only $>$20 hours after the burst when an X-ray
afterglow was found using the {\it BeppoSAX} narrow field instruments \citep{galama}. It seems unlikely, therefore, that we would be able to detect any
observational manifestation of either ionisation or dust destruction by the
GRB jet in this source. {\it Swift} observations of GRBs have revealed later time X-ray
flaring \citep[observed up to 73 hours after the onset of the GRB in rare cases][]{curran} which may also have an effect on the GRB surroundings and at later
times than the GRB jet itself. The X-ray light curve for GRB\,980703 is,
however, not well sampled enough to allow flares to be clearly recognised \citep{gendre1}.

The lower average optical extinction value we have derived for GRB\,980703 remains the
highest of all the extinction values found in the {\it BeppoSAX} GRB afterglow
sample \citep{columnsI}.
In the {\it Swift} era even higher extinction
values have been proposed for so-called `dark' bursts \citep[e.g.][]{jakobsson,rol1} at similar redshifts to
GRB\,980703, such as
GRB\,051022 located at $z=0.8$ in which our line-of-sight to the GRB may pass
through a dusty region of the host galaxy unrelated to the GRB location \citep{rol2}.

\section{Conclusions}
The optical extinction measurements in the literature for GRB\,980703 are
inconsistent, and suggested a decrease in extinction with time indicative of
dust destruction which is predicted to occur in GRBs. We have
re-examined the greatest outlier in these measurements, using broadband SED
fits rather than optical colour, to estimate the extinction. We find a lower
value than previously estimated, in line with most of the later
measurements. We have also investigated the apparent decrease in extinction by
recreating the nIR/optical SEDs at these epochs with extended wavelength
coverage and a wider range of models for the extinction law. 
Both from our new SED fit results and given that the extinction
measurements begin as late as 0.9 days after the burst, we conclude that there
is no evidence of dust destruction by this GRB. 
In the current era of rapid accurate GRB localisations and follow-up we now
have an opportunity to observe some of the effects of GRBs on their
environments, and studies such as this using multiple early-time SEDs will be
of great value.

\begin{acknowledgements}
RLCS acknowledges financial support from STFC and useful discussions with
A.J. van der Horst, P.M. Vreeswijk, K. Wiersema and R.A.M.J. Wijers. We thank the referee for
constructive comments that have improved this manuscript. This research has made use of {\small SAXDAS} linearized and cleaned event
files produced at the {\it BeppoSAX} Science Data Center, and the
software packages {\small ISIS}, {\small Xspec} and {\small IDL}. 
\end{acknowledgements}

\end{document}